\documentclass[11pt,a4paper]{article}
\usepackage[utf8]{inputenc}
\usepackage[margin=2.6cm]{geometry}
\usepackage{amsmath,amssymb}
\usepackage{graphicx}
\usepackage{booktabs}
\usepackage{xcolor}
\usepackage{microtype}
\usepackage{authblk}
\usepackage[colorlinks=true,linkcolor=blue!50!black,citecolor=blue!50!black,urlcolor=blue!50!black]{hyperref}

\newcommand{\KPT}{K_{\mathrm{PT}}}
\newcommand{\Eint}{E_{\mathrm{int}}}
\newcommand{\Fd}{F_d}
\newcommand{\sech}{\operatorname{sech}}

\makeatletter
\renewcommand*{\@fnsymbol}[1]{\ensuremath{\ifcase#1\or\ast\or\dagger\or\ddagger\else\@ctrerr\fi}}
\makeatother

\title{Casimir Attraction and Repulsion in a Magnetic Domain-Wall Background}
\hypersetup{pdftitle={Casimir Attraction and Repulsion in a Magnetic Domain-Wall Background}}

\author[1,2]{In\'es Cavero-Pel\'aez}
\author[1]{Beatriz G\'omez-L\'opez}
\author[3,4]{Jose M. Mu\~noz-Casta\~neda\thanks{Corresponding author: \href{mailto:jose.munoz.castaneda@uva.es}{jose.munoz.castaneda@uva.es}}}
\author[3]{Gonzalo Sancho-Garrido}
\affil[1]{Departamento de F\'isica Te\'orica, Universidad de Zaragoza, 50009 Zaragoza, Spain}
\affil[2]{Centro de Astropart\'iculas y F\'isica de Altas Energ\'ias (CAPA), Universidad de Zaragoza, 50009 Zaragoza, Spain}
\affil[3]{Departamento de F\'isica Te\'orica, At\'omica y \'Optica, Universidad de Valladolid, 47011 Valladolid, Spain}
\affil[4]{IMUVa, Universidad de Valladolid, 47011 Valladolid, Spain}
\date{}

\begin{document}
\maketitle

\begin{abstract}
We study the one-loop Casimir interaction between two $\delta$--$\delta'$ plates in the fixed one-soliton P\"oschl--Teller background of the sine-Gordon kink. The transverse fluctuation problem is the Gaussian truncation of an $S^2$ sigma model, equivalently an easy-axis antiferromagnet with a domain wall, in which the plates exert no classical force on the wall. For $v\ge2|w|$ the fluctuation operators are non-negative, and strictly stable couplings, $v>2|w|$, also exclude zero modes, so the quadratic field theory is consistent at the fixed sine-Gordon mass. The background-dressed relative determinant factorises exactly into effective, gradient-dependent strengths of the two plates, and the sign of each frequency channel of the interaction energy is fixed by their orientation. Two identical, equally oriented, strictly stable plates switch from attraction near the kink to repulsion away from it under a rigid translation at fixed couplings and separation. The interaction energy and the force conjugate to the separation neither share their sign everywhere nor change sign at the same positions: there are regions of negative energy and repulsive force, so attraction and repulsion must be read from the force. For such plates with $w\neq0$ the force is repulsive at short and attractive at large separation, and the interaction energy has a minimum at a separation that depends on the position of the pair. For $v=1.05$, $w=0.5$ and $d=1.5$ the energy changes sign at $c\simeq-0.44$ and $2.63$, and the force at $c\simeq-0.30$ and $2.00$.
\end{abstract}

\section{Introduction}\label{sec:intro}

Quantum vacuum fluctuations generate interactions between neutral objects~\cite{Casimir1948,Milton2001,Bordag2009}. For semitransparent plates modelled by point interactions, these interactions can be computed exactly, and their sign depends on the nature of the boundary conditions: $\delta$--$\delta'$ plates, which are orientation sensitive, can produce attractive, repulsive and vanishing interactions in homogeneous space~\cite{MunozGuilarte2015,MunozGuilarteMoreno2013}.

When the plates are embedded in an inhomogeneous background, the interaction need not depend on their separation alone. In the one-soliton P\"oschl--Teller background of the sine-Gordon kink, the background-dressed relative determinant of two semitransparent $\delta$ plates depends on the individual positions of the plates~\cite{relativeTGTG2026}, and the normalisable zero mode of the background opens an infrared interaction channel that dominates the large-distance behaviour~\cite{zeroModePLB2026}. These results motivate a specific question: can the absolute position of a fixed pair of scatterers change not only the magnitude but also the sign of their Casimir interaction?

We answer this question for $\delta$--$\delta'$ plates in the same background. Although $\delta$--$\delta'$ point interactions are well understood mathematically~\cite{Kurasov1996}, their physical content in Casimir problems has remained implicit. Here we give it explicitly. In a sigma-model realisation, equivalently an easy-axis antiferromagnet with a domain wall~\cite{zeroModePLB2026,Haldane1983,Tveten2013}, a $\delta$--$\delta'$ plate is an anisotropy plane combined with an orientation-sensitive interface that rescales the out-of-plane tilt of the order parameter. For the logarithmic derivative of the fluctuations, the plate acts as a $\delta$ plate whose strength depends on the local gradient of the field. This exact property organises the problem: it gives the stability condition its meaning and fixes the sign of every frequency channel of the interaction energy through the orientation of the plates. The couplings and the separation of the plates are held fixed, and only the position of the pair relative to the kink is varied. Our main result is that strictly stable configurations exist for which both the interaction energy and the force conjugate to the separation change sign under such a rigid translation, and that they do so at different positions. Equally oriented plates necessarily contain one plate whose $\delta'$ coupling points towards its partner. The effective strength of that plate changes sign at a crossover frequency, above which the frequency channels raise the interaction energy. The background shifts this crossover and changes the relative weight of the channels, while the zero mode controls the infrared end, where every channel lowers the energy. Consistently, far from the kink the interaction reduces to the homogeneous massive interaction, whose force is repulsive for suitable couplings and separations, while its formal large-separation asymptotics, governed by the zero mode and supported by direct evaluation of the exact integral, is attractive for all strictly stable couplings.

A second result concerns the notion of attraction itself. If the interaction energy is a power law in the separation, as in scale-free Casimir problems, the energy and the force have the same sign, and the sign of the energy is commonly used to call an interaction attractive or repulsive. In general this identification fails, because the energy need not be monotonic in the separation. Here the energy and the force conjugate to the separation have different signs in whole regions of the $(c,d)$ plane and vanish on different curves. For identical equally oriented plates with non-vanishing $\delta'$ coupling, the force is repulsive at short separation and attractive at large separation, and the interaction energy has a minimum in the separation whose location depends on the position of the pair. Between the zero of the energy and the zero of the force, the energy is negative while the force is repulsive. Attraction and repulsion must therefore be read from the force. This decoupling of the two signs is already present in the homogeneous massive interaction, at very large separation; the background brings it to moderate separations and places it under the control of the position of the pair.

The construction extends that of Ref.~\cite{relativeTGTG2026} to $\delta$--$\delta'$ plates. The one-plate contributions are subtracted in the same background, the reference operator is the sine-Gordon fluctuation operator for every configuration, and the zero mode is retained. The fluctuation mass is fixed by the theory and is the same in all configurations. We show explicitly that, in the regime considered, the plates create no negative-energy states, so that no modification of the mass is required for a consistent quadratic quantum field theory. Plates of this type in the same background were also considered in Ref.~\cite{SantamariaSanz2024}, with a reference problem in which the fluctuation mass depends on the plate configuration.

The paper is organised as follows. Section~\ref{sec:model} defines the fluctuation problem. Section~\ref{sec:sigma} gives its microscopic realisation and the physical meaning of $\delta$--$\delta'$ plates. Section~\ref{sec:stability} establishes stability, Sec.~\ref{sec:det} derives the relative determinant and its infrared structure, and Sec.~\ref{sec:limits} analyses its limits. The sign reversal is established in Sec.~\ref{sec:sign}, and Sec.~\ref{sec:conclusions} summarises the results.

\section{$\delta$--$\delta'$ plates in the P\"oschl--Teller background}\label{sec:model}

We use dimensionless units in which the asymptotic fluctuation mass is one. The transverse fluctuation operator of the sine-Gordon kink is the one-soliton P\"oschl--Teller operator
\begin{equation}
\KPT=-\frac{d^2}{dx^2}+1-2\sech^2x,
\qquad
\sigma(\KPT)=\{0\}\cup[1,\infty),
\label{eq:KPT}
\end{equation}
with normalised zero mode
\begin{equation}
\psi_0(x)=\frac{1}{\sqrt{2}}\sech x,
\qquad
\KPT\psi_0=0.
\label{eq:zeromode}
\end{equation}
The kink is held fixed and centred at $x=0$. The plate positions are parametrised as in Refs.~\cite{relativeTGTG2026,zeroModePLB2026},
\begin{equation}
 a_1=c-\frac{d}{2},\qquad
 a_2=c+\frac{d}{2},\qquad d>0,
\label{eq:geometry}
\end{equation}
so that $d$ is the separation and $c$ the position of the centre of the pair relative to the kink. Since the background is inhomogeneous, a change of $c$ at fixed $d$ and a change of $d$ at fixed $c$ are physically distinct operations.

For Euclidean frequency $\xi>0$ we set $\kappa=\sqrt{1+\xi^2}$ and use the Jost basis
\begin{equation}
 f_+(x)=e^{-\kappa x}(\kappa+\tanh x),\qquad
 f_-(x)=e^{\kappa x}(\kappa-\tanh x),
\label{eq:jost}
\end{equation}
which decays at $+\infty$ and $-\infty$, respectively. With $W[f,g]=fg'-f'g$,
\begin{equation}
W[f_-,f_+]=-2\kappa(\kappa^2-1)=-2\kappa\xi^2 .
\label{eq:wronskian}
\end{equation}
The zero of the Wronskian at $\xi=0$ reflects the normalisable zero mode. At $\kappa=1$ both functions in Eq.~\eqref{eq:jost} reduce to $\sech x$.

A $\delta$--$\delta'$ plate at $x=a_i$ is defined by the Kurasov matching conditions~\cite{Kurasov1996,MunozGuilarte2015}
\begin{equation}
\begin{pmatrix}
\psi(a_i^+)\\
\psi'(a_i^+)
\end{pmatrix}
=C_i
\begin{pmatrix}
\psi(a_i^-)\\
\psi'(a_i^-)
\end{pmatrix},
\qquad
C_i=
\begin{pmatrix}
\alpha_i&0\\
\beta_i&\alpha_i^{-1}
\end{pmatrix},
\label{eq:matching}
\end{equation}
with
\begin{equation}
\alpha_i=\frac{1+w_i/2}{1-w_i/2},
\qquad
\beta_i=\frac{v_i}{1-w_i^2/4},
\qquad |w_i|<2 .
\label{eq:alphabeta}
\end{equation}
For $w_i=0$ the plate reduces to a $\delta$ plate of strength $v_i$. Under the reflection $x\mapsto -x$, with $P=\mathrm{diag}(1,-1)$, one has $PC(v,w)^{-1}P=C(v,-w)$: reflection reverses the orientation of a plate by changing the sign of $w$ while keeping $v$. The couplings $v_i,w_i$ are held fixed when the plates are moved.

Define
\begin{equation}
 H(a,\xi)=
 \begin{pmatrix}
 f_+(a) & f_-(a)\\
 f_+'(a) & f_-'(a)
 \end{pmatrix}.
\label{eq:H}
\end{equation}
The transfer matrix of plate $i$ acting on the coefficients in the Jost basis is
\begin{equation}
M_i(\xi)=H^{-1}(a_i,\xi)\,C_i\,H(a_i,\xi),
\qquad \det M_i=1,
\label{eq:Mi}
\end{equation}
and for the two ordered plates $M_{12}(\xi)=M_2(\xi)M_1(\xi)$.

\section{Microscopic realisation and physical meaning of $\delta$--$\delta'$ plates}\label{sec:sigma}

\subsection{Domain wall and anisotropy planes}

Translating a pair of scatterers relative to a kink is meaningful at one loop only if the scatterers do not exert classical forces on the kink. A generic coupling of the plates to the field supporting the kink generates a tadpole and moves the classical background, and the resulting classical forces would dominate any one-loop effect. Following Ref.~\cite{zeroModePLB2026}, we show that the fluctuation problem of Sec.~\ref{sec:model} is the exact Gaussian truncation of a microscopic model in which no tadpole or classical backreaction arises, and we use this model to identify what a $\delta$--$\delta'$ plate is physically.

Consider an $S^2$ sigma model with field $\mathbf N=(N_x,N_y,N_z)$, $\mathbf N\cdot\mathbf N=4$, and action
\begin{equation}
S_0[\mathbf N]=\frac{1}{2g^2}\int dt\,dx\,\big[\partial_\mu\mathbf N\cdot\partial^\mu\mathbf N-(4-N_z^2)\big].
\label{eq:sigma}
\end{equation}
Related kinks were studied in Refs.~\cite{AlonsoIzquierdo2008,AlonsoIzquierdo2009}. With $\mathbf n=\mathbf N/2$ a unit vector, Eq.~\eqref{eq:sigma} is the long-wavelength theory of an easy-axis antiferromagnet with N\'eel order parameter $\mathbf n$ and easy axis $z$~\cite{Haldane1983,Tveten2013,zeroModePLB2026}. In the coordinates
\begin{equation}
\mathbf N=\Big(\sqrt{4-y^2}\,\sin\tfrac{\phi}{2},\;y,\;\sqrt{4-y^2}\,\cos\tfrac{\phi}{2}\Big),
\end{equation}
the sector $y=0$ is the sine-Gordon model. Its kink $\phi_K=4\arctan e^x$, with
\begin{equation}
\mathbf N_K=(2\sech x,\,0,\,-2\tanh x),
\end{equation}
is a domain wall across which $\mathbf n$ rotates from $+\hat z$ to $-\hat z$ within the $N_xN_z$ plane of spin space. The transverse fluctuation $y=N_y$ tilts $\mathbf n$ out of this rotation plane. Its zero mode, $\psi_0\propto\sech x$, is the rigid rotation of the rotation plane of the wall about the easy axis~\cite{zeroModePLB2026}. The longitudinal fluctuations of $\phi$ contain the translation mode of the wall, which is pinned as in Ref.~\cite{zeroModePLB2026}.

A plate is a plane parallel to the wall, located at $x=a_i$. Its coupling $v_i$ is a hard-axis anisotropy along $N_y$ localised on the plane, $\propto N_y^2$, which penalises the out-of-plane tilt as in Ref.~\cite{zeroModePLB2026}. Its coupling $w_i$ makes the plane orientation sensitive. Formally, the two couplings enter as
\begin{equation}
S_{\mathrm{plates}}\simeq-\frac{1}{2g^2}\int dt\,dx\sum_{i=1}^2\big[v_i\,\delta(x-a_i)+w_i\,\delta'(x-a_i)\big]N_y^2 .
\label{eq:formal}
\end{equation}
The $\delta'$ term is a point dipole layer: it couples to the gradient of $N_y^2$ at the plane, and reflection reverses the sign of $w_i$. Equation~\eqref{eq:formal} is only formal, since the $\delta'$ term requires the value of a field that is discontinuous at the plane, and is therefore ambiguous. We define the plates by their exact self-adjoint completion, the Kurasov interaction~\cite{Kurasov1996,MunozGuilarte2015}. We stress that this is an effective interface and not the zero-thickness limit of an ordinary anisotropy profile. For a non-negative profile $W_\varepsilon$ supported in $|x-a|\le\varepsilon$, the first moment obeys $|\int(x-a)W_\varepsilon\,dx|\le\varepsilon\int W_\varepsilon\,dx$, so a finite $\delta'$ component cannot arise from it at fixed strength. A dipole layer requires anisotropies of both signs, and its point limit depends on how it is taken. At the level of the field, the Kurasov completion is a point defect specified by the junction conditions
\begin{equation}
\phi(a_i^+)=\phi(a_i^-),\qquad y(a_i^+)=\alpha_i\,y(a_i^-),
\label{eq:junction}
\end{equation}
together with the localised term
\begin{equation}
S_{\mathrm{plates}}=-\frac{1}{2g^2}\int dt\,\sum_{i=1}^2\beta_i\,y(t,a_i^-)\,y(t,a_i^+).
\label{eq:Splates}
\end{equation}
Physically, the out-of-plane tilt is rescaled by $\alpha_i$ across the plane, so the plane transmits the tilt asymmetrically, and the plane penalises the product of the tilts on its two sides. For $w_i=0$ there is no junction, $\beta_i=v_i$, and Eq.~\eqref{eq:Splates} reduces to the impurity term $W(x)N_y^2$ of Ref.~\cite{zeroModePLB2026} with $W=\sum_i v_i\delta(x-a_i)$.

\subsection{Absence of tadpoles and Gaussian truncation}

Three properties follow directly.
(i) The wall has $y\equiv0$, so it satisfies Eq.~\eqref{eq:junction} and the plate term and its first variation vanish on it. The wall therefore remains an exact static solution for any plate positions, and its classical energy is independent of $a_1,a_2$: there is no tadpole, no classical force on the plates, and no classical backreaction on the wall. The position dependence of the interaction is thus purely a fluctuation effect.
(ii) Writing $\phi=\phi_K+g\zeta$ and $y=g\eta$, all terms depending on $y$ are even in $y$, so there is no quadratic $\zeta$--$\eta$ mixing. The longitudinal operator is independent of the plates; with the wall pinned, it is identical in the four configurations entering the interaction subtraction and cancels.
(iii) The transverse quadratic energy is
\begin{equation}
\frac12\int_{\mathbb R\setminus\{a_1,a_2\}}dx\,\big[\eta'^2+(1-2\sech^2x)\eta^2\big]
+\frac12\sum_{i}\beta_i\,\eta(a_i^-)\eta(a_i^+),
\qquad \eta(a_i^+)=\alpha_i\eta(a_i^-),
\label{eq:transverseform}
\end{equation}
where the potential term follows from the kinetic term on the wall background exactly as in Ref.~\cite{zeroModePLB2026}. Varying Eq.~\eqref{eq:transverseform} with the junction imposed on the variations gives, at each plane, the natural boundary condition $\eta'(a_i^+)=\beta_i\eta(a_i^-)+\alpha_i^{-1}\eta'(a_i^-)$. Together with the junction, this is precisely the matching condition~\eqref{eq:matching}. The transverse Hessian is therefore $\KPT$ with two $\delta$--$\delta'$ plates, and the complete one-loop relative determinant of the model is the one computed in Sec.~\ref{sec:det}. The interaction studied here is thus a fluctuation-induced interaction between two anisotropy planes, mediated by the out-of-plane fluctuations of a domain wall. For planar plates in $3+1$ dimensions, the wall and the plates extend uniformly in the two parallel directions. The parallel momenta then enter only kinetically, and the same transverse problem arises with the Euclidean frequency replaced by the radial Euclidean momentum (Sec.~\ref{sec:det}).

\subsection{Logarithmic jump: a gradient-dependent $\delta$ plate}

The physical content of the matching condition becomes transparent in terms of the logarithmic derivative of the field, $\ell=\eta'/\eta$, for a solution that does not vanish at the plate. Let $\ell_\mp$ denote its values on the two sides of the plate and $\ell_{\mathrm{av}}=(\ell_++\ell_-)/2$. Equation~\eqref{eq:matching} gives exactly
\begin{equation}
J\equiv\ell_+-\ell_-=\frac{v-2w\ell_-}{(1+w/2)^2}=\frac{v-2w\ell_+}{(1-w/2)^2}=\frac{v-2w\ell_{\mathrm{av}}}{1+w^2/4}.
\label{eq:riccati}
\end{equation}
For a $\delta$ plate of strength $\lambda$ the same jump is $\ell_+-\ell_-=\lambda$. As far as the logarithmic derivative is concerned, a $\delta$--$\delta'$ plate therefore acts as a $\delta$ plate whose strength depends linearly on the local gradient of the field. This is an identity for the jump of a given solution, not an equivalence of operators: the field itself still jumps, $\eta(a^+)=\alpha\eta(a^-)$. To first order in $w$, the symmetric form gives $J=v-2w\ell_{\mathrm{av}}+O(w^2)$, the coupling of a point charge and a point dipole layer suggested by Eq.~\eqref{eq:formal}.

This identity carries the physics of everything that follows. Take $w>0$ for definiteness. A field that decays towards the right ($\ell<0$) sees a stronger plate, and a field that grows towards the right ($\ell>0$) sees a weaker one. If $\ell_{\mathrm{av}}>v/(2w)$ the logarithmic jump is negative, as at an attractive $\delta$ plate, even though $v>0$. The sign of $w$ defines the orientation of the plate, which is weakened by fields whose logarithmic gradient is aligned with it. In the interaction between two plates, each plate is probed by the field that originates at the other. This field decays away from its source, so a plate is weakened when its orientation points towards its partner and strengthened when it points away. We shall say that a plate \emph{faces} its partner in the first case. With this convention, plate 1 faces plate 2 if $w_1>0$, and plate 2 faces plate 1 if $w_2<0$.

\subsection{Limitations}

Two limitations should be stated. First, the junction~\eqref{eq:junction} is defined in the coordinates $(\phi,y)$ and requires $|\alpha_iy|<2$. It defines the plane in a neighbourhood of the wall, which suffices for the semiclassical expansion, but it is not a canonical nonlinear definition of a $\delta'$ plane. As discussed above, it is an effective interface and is not derived from a material anisotropy profile. Second, the classical energy bound used in Ref.~\cite{zeroModePLB2026} relies on the positivity of the impurity term and is not established here. The one-loop calculation requires local stability, i.e.\ non-negativity of the transverse Hessian, which is proven in Sec.~\ref{sec:stability}. As in Ref.~\cite{zeroModePLB2026}, the semiclassical regime in which the rotational zero mode is localised is not analysed here for $\delta$--$\delta'$ plates.

\section{Stability}\label{sec:stability}

The reference operator factorises as $\KPT=A^\dagger A$ with $A=d/dx+\tanh x$. For $\psi$ in the domain of the operator with two plates, integration by parts on each interval and the matching conditions give
\begin{equation}
q[\psi]=\sum_{I}\|A\psi\|_I^2+\sum_{i=1}^2 s_i\,|\psi(a_i^-)|^2,
\qquad
s_i\equiv\alpha_i\beta_i+(\alpha_i^2-1)\tanh a_i=\frac{v_i+2w_i\tanh a_i}{(1-w_i/2)^2},
\label{eq:form}
\end{equation}
where the sum over $I$ runs over the three intervals separated by the plates and $A$ acts on each interval. The surface coefficient $s_i$ depends on the position of the plate because the $\delta'$ coupling makes the plate orientation sensitive. Since $|\tanh a_i|<1$, the condition
\begin{equation}
 v_i\ge2|w_i|\qquad(i=1,2)
\label{eq:stable}
\end{equation}
guarantees $s_i\ge0$ for all finite positions, and hence non-negativity of the two-plate operator and of the two isolated one-plate operators entering the relative determinant. Equation~\eqref{eq:stable} is sufficient; we do not claim that it characterises the stability domain.

The surface coefficient has a direct meaning. Since $\psi_0'/\psi_0=-\tanh x$, Eq.~\eqref{eq:riccati} shows that $s_i=(v_i-2w_i\ell_+)/(1-w_i/2)^2$ with $\ell_+=-\tanh a_i$: it is the logarithmic jump that the plate would impose on a field with the gradient of the zero mode at the plate. Since $|\tanh a_i|<1$ and $v\ge2|w|$ is equivalent to $(v-2w\ell)/(1-w/2)^2\ge0$ for all $|\ell|\le1$, the stability condition ensures that this jump is non-negative wherever the plate is placed.

For strictly stable couplings, $v_i>2|w_i|$, there is moreover no zero mode. If $q[\psi]=0$, then $A\psi=0$ on each interval and $\psi(a_i^-)=0$. On the left interval $A\psi=0$ gives $\psi\propto\sech x$, and $\psi(a_1^-)=0$ forces $\psi=0$ there. The matching condition then gives $\psi(a_1^+)=0$, the same argument applies on the middle interval, and by induction $\psi\equiv0$. The same holds for each isolated plate. Two consequences are used below. The secular factors introduced in Sec.~\ref{sec:det} have no zeros for $\xi>0$, since such a zero would be an eigenvalue $-\xi^2<0$. And the Jost Wronskians at $\xi=0$ do not vanish, since their vanishing would produce a normalisable zero-energy state.

\subsection{Spectrum and consistency of the quadratic field theory}\label{sec:spectrum}

The quadratic quantum field theory around the wall is consistent only if the fluctuation operator has no negative eigenvalues: a negative eigenvalue would produce an imaginary normal-mode frequency, an unbounded quadratic Hamiltonian, and no Fock vacuum. Let $K_{12}$ denote $\KPT$ with the two plates, and $K_i$ the operators with a single plate. The results above show that, for strictly stable couplings and for all plate positions,
\begin{equation}
\sigma(K_{12})=\sigma_{\mathrm{disc}}(K_{12})\cup[1,\infty),\qquad \sigma_{\mathrm{disc}}(K_{12})\subset(0,1),
\label{eq:spectrum}
\end{equation}
and likewise for $K_1$ and $K_2$. The essential spectrum $[1,\infty)$ is not affected by the point interactions, there is no eigenvalue $\le0$, and the discrete eigenvalues lie in the gap. For planar plates in $3+1$ dimensions the normal-mode frequencies are $\omega^2=k_\parallel^2+E$, with $E\in\sigma(K)$ and $k_\parallel$ the momentum parallel to the plates. They are therefore real and positive in all three configurations, and real and non-negative in the reference configuration, where the zero mode of $\KPT$ gives a massless mode propagating along the wall. The quadratic Hamiltonian is bounded below and has a Fock vacuum in each of the four configurations entering the interaction subtraction, all with the same mass, fixed by the sine-Gordon theory. In particular, the contour rotation leading to the Euclidean representation of Sec.~\ref{sec:det} is legitimate, and Eq.~\eqref{eq:EF} already contains the contributions of the bound states in the gap; no separate bound-state sum is to be added.

The physical meaning of this condition is transparent in the realisation of Sec.~\ref{sec:sigma}. A negative eigenvalue of the transverse Hessian would signal that the wall itself is classically unstable against an out-of-plane tilt, and the Gaussian expansion around it would be meaningless. Such an instability cannot be cured by changing the fluctuation mass, which would define a different theory. In the strictly stable domain it does not arise for any positions of the plates.

We have confirmed Eq.~\eqref{eq:spectrum} numerically by two independent methods: the zeros of the Jost function $\mathcal W_{12}(\kappa)$, with $E=1-\kappa^2$, and direct shooting of the fluctuation equation with the matching conditions~\eqref{eq:matching}. The two agree to $10^{-6}$. For the benchmark couplings of Sec.~\ref{sec:sign}, $K_{12}$ has a single bound state for every position examined, with energy
\begin{equation}
\begin{array}{lccccc}
\toprule
c & -3 & -1 & 0 & 1 & 3\\
\midrule
E_{\min} & 0.000998 & 0.267879 & 0.556355 & 0.595759 & 0.028552\\
\bottomrule
\end{array}
\label{eq:Emin}
\end{equation}
The bound state is the zero mode of the wall lifted by the plates. Its energy tends to $0^+$ as $|c|\to\infty$, where the plates leave the wall and the unperturbed zero mode is recovered. Mirror-oriented pairs and single plates with the same couplings also have only positive eigenvalues. Equation~\eqref{eq:stable} is sufficient but not necessary: for $v_i=w_i=0.5$ at $c=0$ and $d=1.5$ the lowest eigenvalue is $0.316>0$. Outside the stable domain negative eigenvalues do occur; for instance, $v_i=-1$, $w_i=0$ at $c=0$ and $d=1.5$ gives $E=-0.701$.

\section{Background-dressed relative determinant}\label{sec:det}

Following Ref.~\cite{relativeTGTG2026}, the interaction energy is defined as $\Eint=E_{12}-E_1-E_2$, with all three vacuum energies computed relative to $\KPT$. With the Jost normalisation of Eq.~\eqref{eq:jost}, the secular factors of the one-plate and two-plate problems are $D_i=(M_i)_{22}$ and $D_{12}=(M_2M_1)_{22}$. A zero of $D_{12}$ at $\xi>0$ signals a solution that decays at both ends, i.e.\ an eigenvalue $-\xi^2$.

Writing $t_i=\tanh a_i$, a direct evaluation of Eq.~\eqref{eq:Mi} gives
\begin{equation}
D_i(\xi)=\frac{\mathcal A_i(\xi)}{2\alpha_i\kappa\xi^2},
\qquad
\mathcal A_i(\xi)=\xi^2P_i(\kappa)+s_i\sech^2a_i,
\qquad
P_i(\kappa)=\kappa(\alpha_i^2+1)+\alpha_i\beta_i,
\label{eq:Di}
\end{equation}
with $s_i$ the surface coefficient of Eq.~\eqref{eq:form}, and
\begin{equation}
(M_i)_{12}=\frac{e^{2\kappa a_i}(\kappa-t_i)}{2\alpha_i\kappa\xi^2}\,\mathcal B_i^-(\xi),
\qquad
(M_i)_{21}=\frac{e^{-2\kappa a_i}(\kappa+t_i)}{2\alpha_i\kappa\xi^2}\,\mathcal B_i^+(\xi),
\label{eq:Moff}
\end{equation}
\begin{equation}
\mathcal B_i^\pm(\xi)=\big[\kappa(\alpha_i^2-1)\pm\alpha_i\beta_i\big](\kappa\pm t_i)-(\alpha_i^2-1)\sech^2a_i .
\label{eq:B}
\end{equation}
Since $(M_2M_1)_{22}=(M_2)_{21}(M_1)_{12}+(M_2)_{22}(M_1)_{22}$, the relative determinant is
\begin{equation}
\Delta(\xi;c,d)=\frac{D_{12}}{D_1D_2}
=1+e^{-2\kappa d}\,
\frac{(\kappa-t_1)(\kappa+t_2)\,
\mathcal B_1^-(\xi)\,\mathcal B_2^+(\xi)}
{\mathcal A_1(\xi)\,\mathcal A_2(\xi)} .
\label{eq:Delta}
\end{equation}
The dependence on $a_1$ and $a_2$ separately, and not only on $d$, is explicit. Equation~\eqref{eq:Delta} has been checked against the direct product of transfer matrices to relative accuracy $10^{-12}$.

\paragraph{Multiple-scattering form.}
Define the Euclidean reflection amplitudes of each plate, alone in the background, for waves incident from the region between the plates:
\begin{equation}
r_1(\xi)=\frac{(M_1)_{12}}{(M_1)_{22}}=e^{2\kappa a_1}(\kappa-t_1)\frac{\mathcal B_1^-}{\mathcal A_1},
\qquad
r_2(\xi)=-\frac{(M_2)_{21}}{(M_2)_{22}}=-e^{-2\kappa a_2}(\kappa+t_2)\frac{\mathcal B_2^+}{\mathcal A_2}.
\label{eq:reflection}
\end{equation}
Then
\begin{equation}
\Delta(\xi;c,d)=1-r_1(\xi)\,r_2(\xi),
\label{eq:TGTG}
\end{equation}
which is the one-dimensional form of the $TGTG$ representation~\cite{KennethKlich2008}, with both reflection amplitudes and the propagation between the plates dressed by the same background~\cite{relativeTGTG2026}. The product $r_1r_2$ is independent of the normalisation of the Jost basis. In a homogeneous background the position dependence of each amplitude factorises into free propagation; here it does not.

\paragraph{Effective strengths and the sign of each channel.}
The field that plate 2 produces in the gap and that probes plate 1 decays towards $-\infty$, i.e.\ it is proportional to $f_-$. Conversely, plate 2 is probed by $f_+$. Their logarithmic gradients at the plates are
\begin{equation}
\ell_1(\xi)=\frac{f_-'(a_1)}{f_-(a_1)}=\kappa-\frac{\sech^2a_1}{\kappa-t_1},
\qquad
\ell_2(\xi)=\frac{f_+'(a_2)}{f_+(a_2)}=-\kappa+\frac{\sech^2a_2}{\kappa+t_2},
\label{eq:ell}
\end{equation}
and we define the effective strengths
\begin{equation}
\Lambda_i(\xi)=\frac{v_i-2w_i\ell_i(\xi)}{(1-w_i/2)^2}=\alpha_i\beta_i-(\alpha_i^2-1)\ell_i(\xi).
\label{eq:Lambda}
\end{equation}
Plate 1 is probed from its right, so $\Lambda_1$ is exactly the logarithmic jump~\eqref{eq:riccati} of a field with gradient $\ell_1$ on that side. Plate 2 is probed from its left, and $\Lambda_2=\alpha_2^2J_2$, with $J_2$ the jump of a field with gradient $\ell_2$ on that side. Since $\alpha_i^2>0$, $\Lambda_i$ and $J_i$ have the same sign. The $\Lambda_i$ are the strengths that enter the reflection amplitudes~\eqref{eq:reflection}: a direct comparison with Eq.~\eqref{eq:B} gives $\mathcal B_1^-=-(\kappa-t_1)\Lambda_1$ and $\mathcal B_2^+=(\kappa+t_2)\Lambda_2$, so that Eq.~\eqref{eq:Delta} takes the exact form
\begin{equation}
\Delta(\xi;c,d)=1-e^{-2\kappa d}\,\frac{(\kappa-t_1)^2(\kappa+t_2)^2}{\mathcal A_1(\xi)\mathcal A_2(\xi)}\,\Lambda_1(\xi)\Lambda_2(\xi).
\label{eq:DeltaLambda}
\end{equation}
This has the structure of the relative determinant of two $\delta$ plates, with strengths replaced by the background-dressed effective strengths of the $\delta$--$\delta'$ plates on the fields exchanged at frequency $\xi$. For strictly stable couplings $\mathcal A_i>0$ (see below), hence
\begin{equation}
\mathrm{sgn}\,\big[\Delta(\xi)-1\big]=-\,\mathrm{sgn}\,\big[\Lambda_1(\xi)\Lambda_2(\xi)\big].
\label{eq:signrule}
\end{equation}
We call a frequency channel energy-lowering (energy-raising) when its contribution to $\Eint$ is negative (positive). A channel is energy-lowering when the two effective strengths have the same sign, and energy-raising when they have opposite signs, as for two $\delta$ plates of strengths of opposite sign. These signs refer to the energy. They do not fix the sign of $\Fd$, since $\partial_d$ also acts on the plate positions, and it is $\Fd$ that decides between attraction and repulsion (Sec.~\ref{sec:EF}). Likewise, a negative $\Lambda_i$ at some Euclidean frequency does not signal an instability; it is the sign of a scattering response.

The gradients~\eqref{eq:ell} are monotonic. $\ell_1$ increases from $-t_1$ at $\xi=0$ to $+\infty$, and $\ell_2$ decreases from $-t_2$ to $-\infty$, because $\partial_\kappa\ell_1=1+\sech^2a_1/(\kappa-t_1)^2>0$ and similarly for $\ell_2$. Combined with $v_i>2|w_i|$, this gives the orientation rule announced in Sec.~\ref{sec:sigma}:
\begin{itemize}
\item A plate that does not face its partner ($w_1\le0$ for plate 1, $w_2\ge0$ for plate 2) has $\Lambda_i\ge(v_i-2|w_i|)/(1-w_i/2)^2>0$ at all frequencies.
\item A plate that faces its partner has $\Lambda_i>0$ at low frequency and $\Lambda_i<0$ at high frequency, with a single crossover. For plate 1 the crossover frequency $\kappa_1^*$ is determined by $\ell_1(\kappa_1^*)=v_1/(2w_1)$, i.e.
\begin{equation}
\kappa_1^*-\frac{\sech^2a_1}{\kappa_1^*-\tanh a_1}=\frac{v_1}{2w_1},
\label{eq:kstar}
\end{equation}
and analogously for plate 2.
\end{itemize}
Three exact consequences follow. If neither plate faces its partner, $w_1\le0\le w_2$, every channel is energy-lowering, $\Delta<1$ for all $\xi$, and $\Eint<0$ for all $c$ and $d$. This includes $\delta$ plates. If exactly one plate faces its partner, channels below the crossover frequency are energy-lowering and channels above it are energy-raising. Equally oriented plates with $w\neq0$ are always in this class. If both plates face each other, the high-frequency channels become energy-lowering again, and energy-raising channels can occur only between the two crossover frequencies. In all cases the channel at $\xi=0$ is energy-lowering: at $\kappa=1$ both gradients reduce to the logarithmic gradient of the zero mode, $\ell_i=-t_i$, and $\Lambda_i(0)=s_i>0$.

\paragraph{Stability and positivity.}
For strictly stable couplings, $D_1$, $D_2$ and $D_{12}$ have no zeros for $\xi>0$ (Sec.~\ref{sec:stability}), and $\Delta\to1$ as $\xi\to\infty$ because of the factor $e^{-2\kappa d}$. Hence $\Delta(\xi)>0$ for all $\xi>0$ and $\log\Delta$ is well defined. Likewise $D_i>0$, since $D_i\to(\alpha_i^2+1)/(2\alpha_i)>0$ as $\xi\to\infty$, and therefore $\mathcal A_i=2\alpha_i\kappa\xi^2D_i>0$.

\paragraph{Infrared behaviour.}
At $\kappa=1$ the factors in Eqs.~\eqref{eq:Di} and~\eqref{eq:B} reduce to
\begin{equation}
\mathcal A_i(0)=s_i\sech^2a_i,\qquad
\mathcal B_i^-(0)=-(1-t_i)\,s_i,\qquad
\mathcal B_i^+(0)=(1+t_i)\,s_i .
\label{eq:IRfactors}
\end{equation}
The infrared limit of each secular factor is thus controlled by the same surface coefficient $s_i$ that controls stability, i.e.\ by the effective strength of the plate on the zero mode. Inserting Eq.~\eqref{eq:IRfactors} into Eq.~\eqref{eq:Delta} and using $(1-t)/(1+t)=e^{-2a}$ gives
\begin{equation}
\Delta(0)=1-e^{-2d}\,\frac{(1-t_1)(1+t_2)}{(1+t_1)(1-t_2)}=0
\label{eq:Delta0}
\end{equation}
exactly. More precisely, let $\mathcal W_I(\kappa)$ denote the Wronskian of the solution of configuration $I\in\{1,2,12\}$ that decays at $-\infty$ with $f_+$. Then $D_I=-\mathcal W_I/(2\kappa\xi^2)$, so that
\begin{equation}
\Delta(\xi)=-2\kappa\xi^2\,\frac{\mathcal W_{12}(\kappa)}{\mathcal W_1(\kappa)\mathcal W_2(\kappa)} .
\label{eq:DeltaW}
\end{equation}
The Wronskians are entire functions of $\kappa$ and, for strictly stable couplings, do not vanish at $\kappa=1$. Therefore
\begin{equation}
\Delta(\xi)=C_{\mathrm{IR}}\,\xi^2+O(\xi^4),\qquad C_{\mathrm{IR}}>0,\qquad
\log\Delta(\xi)=2\log\xi+O(1),\qquad \xi\to0^+ ,
\label{eq:IR}
\end{equation}
where positivity of $C_{\mathrm{IR}}$ follows from $\Delta>0$. The zero mode thus makes $\log\Delta$ logarithmically singular, but integrably so; the interaction energy is infrared finite. For the benchmark of Sec.~\ref{sec:sign} at $c=0$, $C_{\mathrm{IR}}=17.3749$.

\paragraph{Dressed zero-mode couplings.}
Each one-plate secular factor has a Laurent expansion $D_i=\rho_i/\xi^2+\sigma_i+O(\xi^2)$. Following Refs.~\cite{relativeTGTG2026,zeroModePLB2026}, we define the infrared coupling of plate $i$ to the zero-mode fluctuations as the normalisation-independent ratio $\bar\nu_i=\rho_i/\sigma_i$. From Eq.~\eqref{eq:Di},
\begin{equation}
\bar\nu(a;v,w)=\frac{\sech^2a\,(v+2w\tanh a)}{2+v+\tfrac{w^2}{2}-\tfrac12\sech^2a\,(v+2w\tanh a)} .
\label{eq:nubar}
\end{equation}
For $w=0$ this reduces to the dressed coupling $(v/2)\sech^2a/[1+v(\tfrac12-\tfrac14\sech^2a)]$ of a $\delta$ plate~\cite{relativeTGTG2026,zeroModePLB2026}: the numerator is the bare overlap with the zero mode, and the denominator the local response of the massive fluctuations. For strictly stable couplings $\bar\nu_i>0$. The orientation of the plate enters through $v+2w\tanh a$: a plate couples to the zero mode through $v-2w$ deep on the left of the kink and through $v+2w$ deep on the right.

\paragraph{Planar energy and force.}
For planar plates in $3+1$ dimensions, the Euclidean frequency and the two momenta parallel to the plates combine into a radial Euclidean momentum, again denoted by $\xi$, and
\begin{equation}
\frac{\Eint(c,d)}{S}
=\frac{1}{4\pi^2}\int_0^\infty d\xi\,\xi^2\log\Delta(\xi;c,d),
\qquad
\frac{\Fd(c,d)}{S}
=-\left.\frac{\partial}{\partial d}
\frac{\Eint(c,d)}{S}\right|_{c,v_i,w_i}.
\label{eq:EF}
\end{equation}
By Eq.~\eqref{eq:IR} and the factor $e^{-2\kappa d}$, both integrals converge at both ends, and differentiation under the integral is justified. $\Fd$ is the interaction force conjugate to the separation. It is distinct from the full mechanical force on either plate, which also contains derivatives of the position-dependent one-plate self-energies subtracted in $\Eint$~\cite{relativeTGTG2026}. In physical units, lengths are measured in units of the inverse fluctuation mass $m^{-1}$, which is also the width of the kink, and $\Eint/S$ in units of $m^3$.

\section{Limits}\label{sec:limits}

\paragraph{Removal of a plate.}
If $(v_i,w_i)\to(0,0)$, then $\alpha_i\to1$, $\beta_i\to0$ and $\mathcal B_i^\pm\to0$, so $\Delta\to1$ identically and $\Eint=0$, as required of an interaction energy.

\paragraph{$\delta$ plates.}
For $w_1=w_2=0$, Eq.~\eqref{eq:Delta} reduces to
\begin{equation}
\Delta_{\delta\delta}=1-\frac{v_1v_2\,G_{12}^2}{(1+v_1G_{11})(1+v_2G_{22})},
\qquad G_{ij}=G^{\mathrm{PT}}_\xi(a_i,a_j),
\end{equation}
the result of Ref.~\cite{relativeTGTG2026}. Since $G_{ii}>0$, it follows that $0<\Delta_{\delta\delta}\le1$, and hence $\Eint\le0$ for all $c$ and $d$ when $v_1,v_2\ge0$. In this background, a positive interaction energy therefore requires the $\delta'$ coupling.

\paragraph{Homogeneous limit.}
For $|c|\to\infty$ at fixed $d$, $\tanh a_i\to\pm1$ and $\sech^2a_i\to0$, and Eq.~\eqref{eq:Delta} becomes independent of $c$:
\begin{equation}
\Delta_{\mathrm{hom}}(\xi;d)=1-e^{-2\kappa d}\,
\frac{(v_1-2w_1\kappa)(v_2+2w_2\kappa)}
{\big[\kappa(2+\tfrac{w_1^2}{2})+v_1\big]\big[\kappa(2+\tfrac{w_2^2}{2})+v_2\big]} .
\label{eq:Dhom}
\end{equation}
This is the interaction of two $\delta$--$\delta'$ plates in the homogeneous massive vacuum, and the limit is the same on both sides of the kink. Its numerator is the product of the effective strengths~\eqref{eq:Lambda} with the homogeneous gradients $\ell_1=\kappa$ and $\ell_2=-\kappa$. For $w_i=0$ it reduces to the massive double-$\delta$ result $1-v_1v_2e^{-2\kappa d}/[(2\kappa+v_1)(2\kappa+v_2)]$. Two cases matter here. For mirror-oriented plates, $v_1=v_2=v$ and $w_2=-w_1$, the numerator is the square $(v-2w_1\kappa)^2$, so $\Delta_{\mathrm{hom}}=1-e^{-2\kappa d}X$ with $X\ge0$; hence $\log\Delta_{\mathrm{hom}}\le0$ and $\partial_d\log\Delta_{\mathrm{hom}}\ge0$, and neither the homogeneous interaction energy nor the homogeneous force is ever positive. This is the homogeneous instance of the theorem that mirror-symmetric bodies attract~\cite{KennethKlich2006}. Identical, equally oriented plates are not mirror images of each other, and the theorem does not apply. For them the numerator is $v^2-4w^2\kappa^2$, and the effective strength of the facing plate is negative above $\kappa=v/(2|w|)$, which is Eq.~\eqref{eq:kstar} with $\sech^2a_1=0$. When $v$ is close to $2|w|$, almost the whole frequency range raises the energy. For the benchmark couplings of Sec.~\ref{sec:sign} the homogeneous energy and force are then positive up to very large separations; they change sign only at $d\simeq15.18$ and $d\simeq15.65$, respectively.

\paragraph{Large separation: the zero-mode channel.}
Let $d\to\infty$ at fixed $c$. The plates then lie in the tails of the zero mode, $t_1\to-1$, $t_2\to1$ and $\sech^2a_{1,2}\simeq4e^{\pm2c-d}$. On the infrared scale $\xi^2=O(e^{-d})$ one has $\kappa=1+O(e^{-d})$, so that $\mathcal A_i\simeq P_i(1)\,(\xi^2+\gamma_ie^{-d})$ and, from Eq.~\eqref{eq:IRfactors}, $\mathcal B_1^-\simeq-2s_1$ and $\mathcal B_2^+\simeq2s_2$, with
\begin{equation}
\gamma_1=\frac{4e^{2c}(v_1-2w_1)}{2+v_1+w_1^2/2},
\qquad
\gamma_2=\frac{4e^{-2c}(v_2+2w_2)}{2+v_2+w_2^2/2},
\label{eq:gamma}
\end{equation}
which are the leading terms of the dressed couplings, $\bar\nu_i=\gamma_ie^{-d}[1+O(e^{-d})]$. Setting $\xi=e^{-d/2}y$, Eq.~\eqref{eq:Delta} gives the pointwise limit
\begin{equation}
\Delta(e^{-d/2}y)\longrightarrow\frac{y^2(y^2+\gamma_1+\gamma_2)}{(y^2+\gamma_1)(y^2+\gamma_2)},\qquad d\to\infty,
\label{eq:Dinf}
\end{equation}
which has the zero-mode relative-determinant structure found for $\delta$ plates~\cite{relativeTGTG2026,zeroModePLB2026}. Taking this limit formally through the integral~\eqref{eq:EF} gives
\begin{equation}
\frac{\Eint}{S}=\frac{e^{-3d/2}}{12\pi}\,\Gamma+o(e^{-3d/2}),
\qquad
\frac{\Fd}{S}=\frac{e^{-3d/2}}{8\pi}\,\Gamma+o(e^{-3d/2}),
\qquad
\Gamma=\gamma_1^{3/2}+\gamma_2^{3/2}-(\gamma_1+\gamma_2)^{3/2}.
\label{eq:largeD}
\end{equation}
For strictly stable couplings $\gamma_1,\gamma_2>0$, and since $x^{3/2}$ is strictly superadditive, $\Gamma<0$: in the formal leading term both the energy and the force are negative for all strictly stable couplings. It decays as $e^{-3d/2}$, more slowly than the homogeneous massive interaction, whose decay is governed by $e^{-2d}$. Near the edge of the sufficient stability condition, $v_1-2w_1$ or $v_2+2w_2$ becomes small and the corresponding plate decouples from the zero-mode channel at leading order.

For $\delta$ plates, the interchange of the limit~\eqref{eq:Dinf} with the integral was justified in Ref.~\cite{relativeTGTG2026} by a $d$-independent integrable majorant. We have not adapted that bound to $\delta$--$\delta'$ plates, and the asymptotic form of $\Fd$ requires in addition control of the $d$-derivative of the remainder. Equation~\eqref{eq:largeD} is therefore a formal asymptotic result, supported by the pointwise limit and by the direct comparison with the exact integrals in Table~\ref{tab:largeD}. None of the results of Sec.~\ref{sec:sign} depends on it.

\begin{table}[htb]
\centering
\begin{tabular}{ccccc}
\toprule
 & \multicolumn{2}{c}{$c=0$} & \multicolumn{2}{c}{$c=1$}\\
\cmidrule(lr){2-3}\cmidrule(lr){4-5}
$d$ & $\Eint/\Eint^{\mathrm{lead}}$ & $\Fd/\Fd^{\mathrm{lead}}$ & $\Eint/\Eint^{\mathrm{lead}}$ & $\Fd/\Fd^{\mathrm{lead}}$\\
\midrule
6  & 0.7143 & 0.6613 & 1.0964 & 1.2007\\
10 & 0.9405 & 0.9225 & 0.9687 & 0.9618\\
14 & 0.9909 & 0.9880 & 0.9942 & 0.9924\\
18 & 0.9987 & 0.9982 & 0.9991 & 0.9989\\
\bottomrule
\end{tabular}
\caption{Ratio of the exact planar interaction energy and force, Eq.~\eqref{eq:EF}, to the leading large-separation terms of Eq.~\eqref{eq:largeD}, for identical equally oriented plates with $v=1.05$ and $w=0.5$.}
\label{tab:largeD}
\end{table}

\section{Position-controlled sign reversal}\label{sec:sign}

The two limits of Sec.~\ref{sec:limits} point to opposite signs for suitable couplings. The formal large-separation asymptotics, supported by direct evaluation of the exact integral, is attractive for every strictly stable pair, whereas at moderate separations the homogeneous force between identical equally oriented plates is repulsive when $v$ is close to $2|w|$. We now show, using only the exact relative determinant, that at a fixed, moderate separation the interaction changes sign under rigid translation.

\subsection{Benchmark}

\begin{figure}[htb]
\centering
\includegraphics[width=0.62\linewidth]{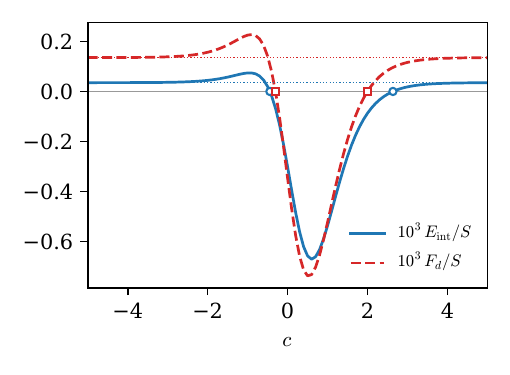}
\caption{Interaction energy and force conjugate to the separation for two identical, equally oriented $\delta$--$\delta'$ plates with $v=1.05$, $w=0.5$ and $d=1.5$, as functions of the position $c$ of the pair relative to the kink. Couplings and separation are fixed. Open symbols mark the zeros, and dotted lines the homogeneous values approached as $|c|\to\infty$.}
\label{fig:same}
\end{figure}

Consider two identical plates with the same orientation, $v_1=v_2=v$ and $w_1=w_2=w$, and choose
\begin{equation}
 v=1.05,\qquad w=0.5,\qquad d=1.5 .
\label{eq:benchmark}
\end{equation}
Equation~\eqref{eq:stable} holds strictly, so by Sec.~\ref{sec:stability} every configuration is free of negative and zero modes for all $c$. The numerical evaluation of Eqs.~\eqref{eq:Delta} and~\eqref{eq:EF} gives
\begin{equation}
\begin{array}{lcc}
\toprule
 & \Eint/S & \Fd/S\\
\midrule
 c=0 & -3.01298\times10^{-4} & -3.46610\times10^{-4}\\
 c=3 & +1.83993\times10^{-5} & +1.16236\times10^{-4}\\
 |c|\to\infty & +3.51901\times10^{-5} & +1.35435\times10^{-4}\\
\bottomrule
\end{array}
\label{eq:benchmarktable}
\end{equation}
The same pair of plates, at the same separation and with the same couplings, is therefore attractive near the kink ($\Fd<0$) and repulsive after a rigid translation away from it ($\Fd>0$). Both $\Eint$ and $\Fd$ are continuous in $c$, so each has at least one zero in between. Within $-5\le c\le5$ they have two zeros each: $\Eint$ vanishes at $c\simeq-0.4432$ and $c\simeq2.6344$, and $\Fd$ at $c\simeq-0.3038$ and $c\simeq1.9971$. The zeros of $\Eint$ and $\Fd$ do not coincide: between them the energy is negative while the force is repulsive. This is analysed in Sec.~\ref{sec:EF}. The curves are shown in Fig.~\ref{fig:same}. The interaction is not symmetric under $c\mapsto-c$, because reflection maps equally oriented plates with coupling $w$ to equally oriented plates with coupling $-w$.

\begin{figure}[htb]
\centering
\includegraphics[width=0.62\linewidth]{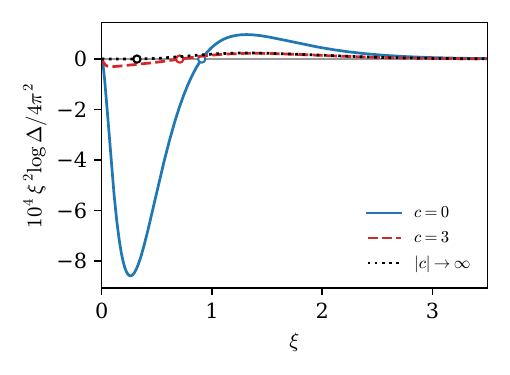}
\caption{Frequency-resolved integrand of the interaction energy, Eq.~\eqref{eq:EF}, for the benchmark at $c=0$, $c=3$ and in the homogeneous limit. Open circles mark the crossover frequency of Eq.~\eqref{eq:kstar}, below which channels lower the energy and above which they raise it. Near the kink the soft channels dominate.}
\label{fig:freq}
\end{figure}

\paragraph{Frequency-resolved picture.}
By Eq.~\eqref{eq:signrule}, for the benchmark the facing plate is plate 1, and channels with $\kappa<\kappa_1^*$ lower the energy while channels with $\kappa>\kappa_1^*$ raise it. The position of the pair acts in two ways. First, the kink raises the crossover frequency through the term $\sech^2a_1/(\kappa-\tanh a_1)$ in Eq.~\eqref{eq:kstar}. Second, the background changes the relative weight of the channels: near the kink the soft channels carry most of the interaction, and their infrared end is controlled by the zero mode. Splitting the integral~\eqref{eq:EF} exactly at $\xi^*=\sqrt{\kappa_1^{*2}-1}$ gives Table~\ref{tab:split} and Fig.~\ref{fig:freq}. Near the kink the energy-lowering soft channels outweigh the energy-raising hard channels by a factor of four. Away from it they are suppressed, the crossover tends to its homogeneous value $\kappa=v/(2w)=1.05$, and the energy-raising channels win.

\begin{table}[htb]
\centering
\begin{tabular}{cccccc}
\toprule
$c$ & $\kappa_1^*$ & $\xi^*$ & lowering ($\xi<\xi^*$) & raising ($\xi>\xi^*$) & $\Eint/S$\\
\midrule
0 & 1.3505 & 0.9076 & $-4.023\times10^{-4}$ & $+1.010\times10^{-4}$ & $-3.013\times10^{-4}$\\
1 & 1.6972 & 1.3714 & $-5.598\times10^{-4}$ & $+2.376\times10^{-5}$ & $-5.360\times10^{-4}$\\
2 & 1.4882 & 1.1022 & $-1.092\times10^{-4}$ & $+2.250\times10^{-5}$ & $-8.665\times10^{-5}$\\
3 & 1.2256 & 0.7085 & $-1.239\times10^{-5}$ & $+3.079\times10^{-5}$ & $+1.840\times10^{-5}$\\
$|c|\to\infty$ & 1.0500 & 0.3202 & $-5.1\times10^{-8}$ & $+3.524\times10^{-5}$ & $+3.519\times10^{-5}$\\
\bottomrule
\end{tabular}
\caption{Exact splitting of the interaction energy of the benchmark into the energy-lowering channels below and the energy-raising channels above the crossover frequency of Eq.~\eqref{eq:kstar}.}
\label{tab:split}
\end{table}

\paragraph{Robustness.}
On the strictly stable domain, the integrands in Eq.~\eqref{eq:EF} and their $d$-derivatives are continuous in $(c,d,v_i,w_i)$ and admit locally uniform integrable majorants: near $\xi=0$ this follows from Eq.~\eqref{eq:IR}, with $C_{\mathrm{IR}}$ continuous and positive, and at large $\xi$ from the factor $e^{-2\kappa d}$. Hence $\Eint$ and $\Fd$ are continuous functions of all parameters, and the strict inequalities $\Fd(0)<0<\Fd(3)$ persist on an open neighbourhood of the benchmark in parameter space. The sign reversal is therefore not a fine-tuned property of the benchmark.

\paragraph{Interpretation.}
The sign of every frequency channel is fixed exactly by the orientation of the plates, Eq.~\eqref{eq:signrule}. The sign of the integrated interaction is a balance between energy-lowering soft channels and energy-raising hard channels, and the position of the pair relative to the kink controls that balance. This picture is consistent with the limits of Sec.~\ref{sec:limits}: as the pair leaves the kink the interaction tends to the homogeneous one, here with positive energy and repulsive force, and its formal large-separation asymptotics is attractive. At $d=1.5$ the balance is not given by an asymptotic expansion; it is computed exactly, as in Table~\ref{tab:split}. We do not attribute the reversal at this separation to the zero mode alone: the low-frequency channels are not a projection onto the zero mode, and they include the response of the massive continuum dressed by the background.

\paragraph{Numerical evaluation.}
The integrals in Eq.~\eqref{eq:EF} were evaluated by adaptive quadrature. The force was obtained by fourth-order finite differences in $d$, and the zeros by bracketing root finding. Equation~\eqref{eq:Delta} agrees with the direct product of transfer matrices to $10^{-12}$, and the homogeneous values in Eq.~\eqref{eq:benchmarktable} agree with Eq.~\eqref{eq:Dhom}. The values at $c=0$ and $c=3$ were also obtained in an independent calculation by two of the authors.

\subsection{Energy and force: attraction is decided by the force}\label{sec:EF}

The benchmark shows that $\Eint$ and $\Fd$ do not vanish at the same positions. For $-0.4432<c<-0.3038$ and for $1.9971<c<2.6344$ the interaction energy is negative while the force is repulsive. Throughout $-0.4432<c<2.6344$ the energy keeps its sign while the force changes sign twice. The sign of the interaction energy is therefore not a criterion for attraction. We call the interaction attractive when $\Fd<0$ and repulsive when $\Fd>0$.

The identification of the two signs is justified when the energy is a power law. If $\Eint=A\,d^{-n}$ with $n>0$ and $A$ independent of $d$, then $\Fd=n\Eint/d$ and the signs coincide, as for ideal boundary conditions on massless fields. The converse does not hold, and the presence of scales does not by itself separate the two signs. What separates them is that $\Eint$ is not monotonic in $d$. This already happens in the homogeneous massive limit, Eq.~\eqref{eq:Dhom}: for the benchmark couplings the homogeneous energy vanishes at $d\simeq15.18$ and the homogeneous force at $d\simeq15.65$, and between them the energy is negative while the force is repulsive. The inhomogeneous background is therefore not necessary for the decoupling of the two signs. What it adds is geometric control: the zeros depend on the position $c$ of the pair and are brought to moderate separations.

\paragraph{Short and large separations.}
For $d\to0$ at fixed $c$ the integral~\eqref{eq:EF} is dominated by $\xi\sim1/d$. There $\Lambda_1\simeq-(\alpha_1^2-1)\kappa$, $\Lambda_2\simeq(\alpha_2^2-1)\kappa$ and $\mathcal A_i\simeq\xi^2\kappa(\alpha_i^2+1)$, so that $\Delta\simeq1+R\,e^{-2\xi d}$ and
\begin{equation}
\frac{\Eint}{S}\simeq-\frac{\mathrm{Li}_4(-R)}{16\pi^2d^3},\qquad
\frac{\Fd}{S}\simeq-\frac{3\,\mathrm{Li}_4(-R)}{16\pi^2d^4},\qquad
R=\frac{w_1w_2}{(1+w_1^2/4)(1+w_2^2/4)} .
\label{eq:short}
\end{equation}
This is a scale-free regime: it depends neither on $v_i$ nor on $c$, and the energy and the force have the same sign. For $w_1w_2>0$, in particular for equally oriented plates with $w\neq0$, $R>0$ and both are positive, so the force is repulsive at short separation. For $w_1w_2=0$ the leading term vanishes and Eq.~\eqref{eq:short} does not apply; for $\delta$ plates with $v>0$ the short-distance energy and force are negative. For the benchmark, $R=0.2215$, and the ratio of the exact energy at $c=0$ to Eq.~\eqref{eq:short} is $0.80$, $0.90$, $0.95$ and $0.98$ at $d=0.2$, $0.1$, $0.05$ and $0.02$. Equation~\eqref{eq:short} is the leading term of the energy; for the force, a complete proof would also control the $d$-derivative of the remainder, and the ratio of the exact force to Eq.~\eqref{eq:short} is $0.87$, $0.94$, $0.97$ and $0.99$ at the same separations. At large separation, by the formal asymptotics~\eqref{eq:largeD}, both the energy and the force are negative.

\paragraph{Zeros and minimum of the energy.}
For identical equally oriented plates with $w\neq0$ the energy is thus positive at short separation and negative at large separation. At fixed $c$ it therefore vanishes at some separation $d_E(c)$ and has a negative minimum at some $d_F(c)>d_E(c)$, where $\Fd=0$. In all positions examined, $-4\le c\le4$, each of $\Eint$ and $\Fd$ has exactly one zero in $d$. This uniqueness is a numerical observation for the benchmark couplings, not a theorem. In the domain examined there are three regimes:
\begin{equation}
\begin{array}{lll}
d<d_E(c): & \Eint>0, & \Fd>0\ \text{(repulsive)},\\
d_E(c)<d<d_F(c): & \Eint<0, & \Fd>0\ \text{(repulsive)},\\
d>d_F(c): & \Eint<0, & \Fd<0\ \text{(attractive)}.
\end{array}
\end{equation}
In the intermediate regime the energy is negative and the force is repulsive. The opposite combination, a positive energy with an attractive force, does not occur for these plates in the region examined. Since $\Fd>0$ below $d_F$ and $\Fd<0$ above it, $d_F(c)$ is a local minimum of the irreducible interaction energy with respect to the separation at fixed centre position; for instance, $\partial_d^2(\Eint/S)\simeq2.87\times10^{-3}>0$ at $c=0$. It is not a mechanical equilibrium of the pair: at the same point $-\partial_c(\Eint/S)\simeq1.45\times10^{-3}\neq0$, and the total mechanical forces also contain the position-dependent one-plate self-energies subtracted in $\Eint$.

\begin{figure}[htb]
\centering
\includegraphics[width=0.98\linewidth]{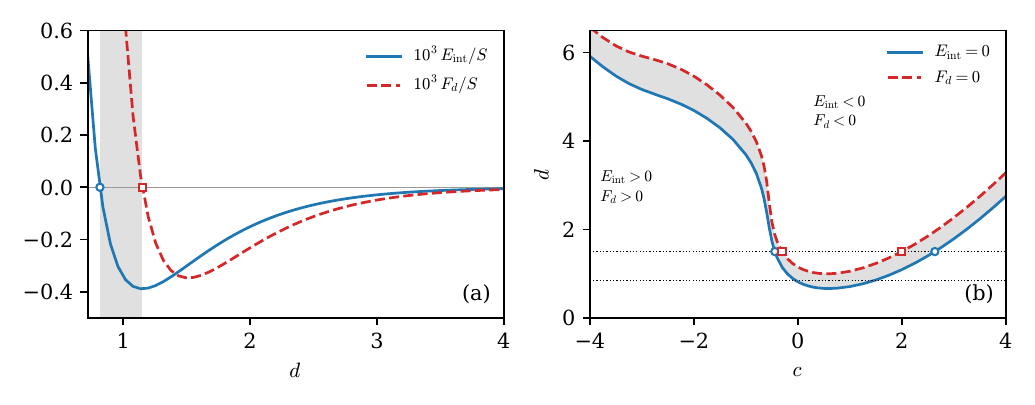}
\caption{Energy versus force for identical, equally oriented plates with $v=1.05$ and $w=0.5$. (a) Interaction energy and force as functions of the separation at $c=0$. In the shaded interval, $d_E<d<d_F$, the energy is negative and the force is repulsive. (b) Curves $\Eint=0$ (solid) and $\Fd=0$ (dashed) in the $(c,d)$ plane. In the shaded band the energy is negative and the force is repulsive. The dotted lines are the cuts $d=1.5$, with the four zeros of Fig.~\ref{fig:same}, and $d=0.85$.}
\label{fig:EF}
\end{figure}

Figure~\ref{fig:EF}(a) shows the case $c=0$, where $d_E=0.8164$ and $d_F=1.1512$. Figure~\ref{fig:EF}(b) shows the curves $d_E(c)$ and $d_F(c)$, which divide the $(c,d)$ plane into the three regimes. Both curves have their minimum near $c\simeq0.6$, where $d_E=0.664$ and $d_F=0.997$. They rise steeply on the left of the kink and more slowly on its right. In the homogeneous interaction the two zeros are at $d\simeq15.18$ and $d\simeq15.65$.

\paragraph{Rigid translations.}
A rigid translation at fixed $d$ is a horizontal line in Fig.~\ref{fig:EF}(b). The benchmark line $d=1.5$ crosses the curve $\Fd=0$ at $c\simeq-0.3038$ and $1.9971$, and the curve $\Eint=0$ at $c\simeq-0.4432$ and $2.6344$. The position-controlled sign reversal is thus the statement that the separation $d_F(c)$ of the minimum of the interaction energy depends on the position of the pair. A line with $0.664<d<0.997$ crosses only the curve $\Eint=0$. For $d=0.85$ the energy is negative for $-0.048<c<1.481$ and positive otherwise, while the force is repulsive at every position examined, $-6\le c\le8$. Hence the energy can change sign under a translation while the force does not, and, within $-0.4432<c<2.6344$ at $d=1.5$, the force changes sign while the energy does not.

\subsection{Mirror-oriented plates: a parity check}

\begin{figure}[htb]
\centering
\includegraphics[width=0.62\linewidth]{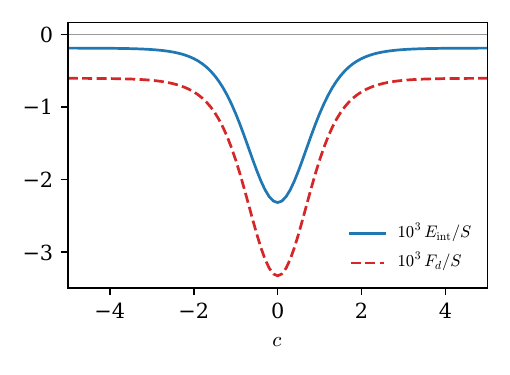}
\caption{Parity check: mirror-oriented plates with $v_1=v_2=1.05$, $w_1=-0.5$, $w_2=+0.5$ and $d=1.5$. The curves are even in $c$, as required by parity, and the force is attractive throughout the displayed interval.}
\label{fig:mirror}
\end{figure}

By the reflection property of Sec.~\ref{sec:model} and the evenness of the background,
\begin{equation}
\Delta(\xi;c,d;v_1,w_1,v_2,w_2)=\Delta(\xi;-c,d;v_2,-w_2,v_1,-w_1).
\end{equation}
For mirror-oriented plates, $v_1=v_2$ and $w_2=-w_1$, the interaction is therefore even in $c$. Figure~\ref{fig:mirror} shows the case $w_1=-0.5$, $w_2=+0.5$ with the same $v$ and $d$. The numerical curves are even in $c$ within numerical accuracy, and the force is attractive over the displayed interval. In this configuration neither plate faces its partner, so by Sec.~\ref{sec:det} every channel is energy-lowering and $\Eint<0$ for all $c$ and $d$; the attraction of the force is observed numerically. The opposite choice, $w_1=+0.5$ and $w_2=-0.5$, is also even in $c$, but there both plates face each other, and no sign-definiteness follows at finite $c$.

\section{Conclusions}\label{sec:conclusions}

We have computed and interpreted the one-loop Casimir interaction between two $\delta$--$\delta'$ plates in the fixed one-soliton P\"oschl--Teller background. The transverse problem is the Gaussian truncation of an $S^2$ sigma model, equivalently an easy-axis antiferromagnet with a domain wall. In this effective realisation, the plates are anisotropy planes with an orientation-sensitive interface that rescales the out-of-plane tilt. They exert no classical force on the wall, so their position dependence is purely a fluctuation effect. Exactly, the logarithmic derivative of any field jumps across a $\delta$--$\delta'$ plate as at a $\delta$ plate whose strength depends on the local gradient. The condition $v_i\ge2|w_i|$ ensures non-negativity of the fluctuation operators, while strictly stable couplings, $v_i>2|w_i|$, also exclude zero modes at all finite plate positions. The quadratic quantum field theory is then consistent at the fixed sine-Gordon mass, and no adjustment of the mass is needed. The jump identity gives this condition a direct meaning, the non-negativity of the jump on the zero mode at every position, and it makes the coupling to the zero mode depend on the side of the wall on which the plate sits.

The background-dressed relative determinant factorises exactly into the effective strengths of the two plates. The sign of each frequency channel is therefore fixed by orientation. Plates that do not face each other have a negative interaction energy for all positions and separations. The effective strength of a facing plate changes sign at a single crossover frequency, above which the channels raise the energy. Equally oriented plates always contain one facing plate. The background shifts the crossover frequency and the weight of the channels, and the infrared end, controlled by the zero mode, always lowers the energy. Far from the kink the interaction tends to the homogeneous massive interaction, whose force is repulsive for suitable couplings and separations, while its formal large-separation asymptotics is attractive and decays as $e^{-3d/2}$.

As a consequence, the sign of the interaction is controlled by the position of the pair. For $v=1.05$, $w=0.5$ and $d=1.5$, both the interaction energy and the separation force change sign under a rigid translation relative to the kink, at fixed couplings and separation, and this persists on an open set of parameters.

The energy and the force do not share their sign. For identical equally oriented plates with $w\neq0$ the force is repulsive at short separation and attractive at large separation, and it vanishes at a separation $d_F(c)$, a minimum of the interaction energy in $d$ at fixed $c$, that depends on the position of the pair. The energy vanishes on a different curve, $d_E(c)<d_F(c)$, and between the two the energy is negative while the force is repulsive. Under a rigid translation the energy can change sign while the force does not, and conversely. The sign of the interaction energy therefore does not decide between attraction and repulsion; the force does. This decoupling is already present in the homogeneous massive interaction at very large separation. The background brings it to moderate separations and makes it controllable by the position of the pair. The zero curves have been determined numerically for the benchmark couplings; their uniqueness is not proven. A fixed inhomogeneous background can thus act as a geometric control parameter for the sign of a Casimir interaction between otherwise unchanged objects.

Several questions remain open: the complete stability domain of $\delta$--$\delta'$ plates in this background, a uniform majorant that would turn the large-separation asymptotics into a theorem, the semiclassical domain of validity of the sigma-model realisation, and the total mechanical forces, which require the renormalised position-dependent one-plate self-energies.

\section*{Acknowledgements}
J.M.M.-C. and G.S.-G. acknowledge partial financial support from Grant PID2021-123251NB-I00 funded by MCIN/AEI/10.13039/501100011033 and by the European Union, and from the Department of Education of the Junta de Castilla y Le\'on and FEDER Funds (Reference No.\ CLU-2025-1-02-IMUVA). G.S.-G. also acknowledges financial support from a University of Valladolid predoctoral contract co-funded by Banco Santander (Grant No.\ CONTPR-2024-310) and from the Formaci\'on de Profesorado Universitario (FPU) programme of the Spanish Ministry of Science, Innovation and Universities (Grant No.\ FPU24/01715). I.C.-P. acknowledges partial financial support from Spanish MINECO/FEDER Grant PID2024-160228NB-I00, funded by MCIN/AEI/10.13039/501100011033 and the European Regional Development Fund (ERDF, ``A way of making Europe''), and by Grant E21-26R funded by the Government of Arag\'on and the European Union.

\section*{Data and code availability}
No external datasets were used. All numerical results follow from the explicit formulas of this paper.

\end{document}